\documentclass[12pt]{article}

\newcommand{\bbr}{I\!\! R}

\newcommand{\x}{arXiv:}
\newcommand{\m}{\mathrm}

\usepackage{graphicx}

\begin{document}
\thispagestyle{empty}
\begin{center}

\null \vskip-1truecm \vskip2truecm

{\Large{\bf \textsf{Decoupling Inflation From the String Scale
}}}

\bigskip

{\Large{\bf \textsf{}}}

{\large{\bf \textsf{}}}

\vskip1truecm

{\large \textsf{Brett McInnes}}

\vskip1truecm

\textsf{\\  National
  University of Singapore}

\textsf{email: matmcinn@nus.edu.sg}\\

\end{center}
\vskip1truecm \centerline{\textsf{ABSTRACT}} \baselineskip=15pt
\medskip

When Inflation is embedded in a fundamental theory, such as string theory, it typically begins when the Universe is already substantially larger than the fundamental scale [such as the one defined by the string length scale]. This is naturally explained by postulating a
pre-inflationary era, during which the size of the Universe grew from the fundamental scale to the initial inflationary scale. The problem then arises of maintaining the [presumed] initial spatial homogeneity
throughout this era, so that, when it terminates, Inflation is able to begin in its potential-dominated state. Linde \cite{kn:lindetypical}
has proposed that a spacetime with compact negatively curved spatial sections can achieve this,
by means of chaotic mixing. Such a compactification will however lead to a Casimir energy, which can lead to effects that defeat the purpose unless the coupling to gravity is suppressed. We estimate the value of
this coupling required by the proposal, and use it to show that the pre-inflationary spacetime is stable, despite the violation of the Null Energy Condition entailed by the Casimir energy.

\newpage

\addtocounter{section}{1}
\section* {\large{\textsf{1. Getting Inflation Started}}}
It has long been understood [see for example \cite{kn:albrecht}\cite{kn:dyson} and their references]
that, at the beginning of Inflation \cite{kn:lindenew}, the inflaton must have been in an extremely non-generic state. This state is associated with the extreme homogeneity of the relevant region of three-dimensional space. Understanding the ultimate origin of this homogeneity is of course a basic objective of theories of the ``Arrow of Time" \cite{kn:carroll}\cite{kn:wald}\cite{kn:arrow}\cite{kn:mersini}\cite{kn:bojo}\cite{kn:greene}. Here, however, we shall not be concerned with that deep question; instead we shall focus on a less profound but equally vital problem: granted that, for some reason yet to be fully understood, the spatial geometry was very homogeneous at the earliest times, \emph{how was that extreme homogeneity maintained during the era before Inflation began?}

The question arises because, in many inflationary models ---$\,$ particularly those which embed Inflation in string theory \cite{kn:racetrack}\cite{kn:liameva}\cite{kn:baumann} ---$\,$ the length scale of the Universe at the inception of Inflation is substantially larger [say, by three or four orders of magnitude] than the characteristic length scale of a fundamental theory, such as the string length. It is natural to postulate that there must have been a ``pre-inflationary" era during which the Universe grew [from the string length scale] at a relatively modest pace, until Inflation was ready to start. But we then have to explain how, throughout that period, despite the natural tendency of inhomogeneities to grow under the action of strong gravitational fields, the extreme initial homogeneity was maintained, so that the inflaton remained in its potential-dominated [almost perfectly homogeneous] state. To put it another way, the entropy associated with the gravitational degrees of freedom increased only to a negligible extent during this era. Although this does not violate the second law of thermodynamics, it does not seem plausible. [The question as to whether effects arising during this era can be observed is surveyed in \cite{kn:amelia}; see for example \cite{kn:pre} for more recent developments.]

This interesting problem was raised by Linde \cite{kn:lindetypical}, who proposed an ingenious solution to it [which we shall re-formulate in the language of conformal time and Penrose diagrams]. As is well known, the global version of [simply connected] de Sitter spacetime, the version with spherical spatial sections, has a relatively short conformal lifetime: the Penrose diagram is only as high as it is wide. However, parts of de Sitter spacetime can be foliated by spatial sections which are flat or negatively curved. Regarded as spacetimes in their own right, these have infinite conformal lifetimes, as we shall review later. Now Linde proposes that we take these non-spherical spatial sections and \emph{compactify} them by taking a topological quotient. The resulting spacetime is to be used as a model of the whole era from the beginning of time to the end of Inflation. For example, if we begin with the version of de Sitter spacetime which is foliated by copies of the three-dimensional hyperbolic space H$^3$, then we can obtain compact spatial sections of the form H$^3$/$\Gamma$, where $\Gamma$ is one
of the many infinite discrete subgroups \cite{kn:thurston} of O(1,3) that can act
freely, isometrically, and properly discontinuously on H$^3$. The Penrose diagram of the corresponding spacetime then has a finite width, but an infinite height\footnote{In reality, one might wish to cut off the part of the spacetime where the scale factor approaches zero, if for example one considers that string T-duality renders arbitrarily short scales unphysical, or in theories of creation from ``nothing". The reader may therefore prefer to regard the height of the diagram, at this point of the argument, as finite but large [compared to its width].} [measured downwards from future conformal infinity].

This has an important consequence: it means that, during the earliest times, signals sent outward from a given point can ``wind around the Universe" and return to a neighbourhood of that point. Note that this is not the case in simply connected global de Sitter spacetime, since, as observed above, its Penrose diagram is square. In the case of the spacetime we have constructed here, we can identify the top square of the Penrose diagram, where no signal sent outward even begins to return, with the usual square de Sitter diagram which can describe the inflationary era. The pre-inflationary era then corresponds to the remaining [lower] part of the diagram, where global winding is possible.
The Penrose diagram representing these ideas is given in Figure 1.

\begin{figure}[!h]
\centering
\includegraphics[width=0.6\textwidth]{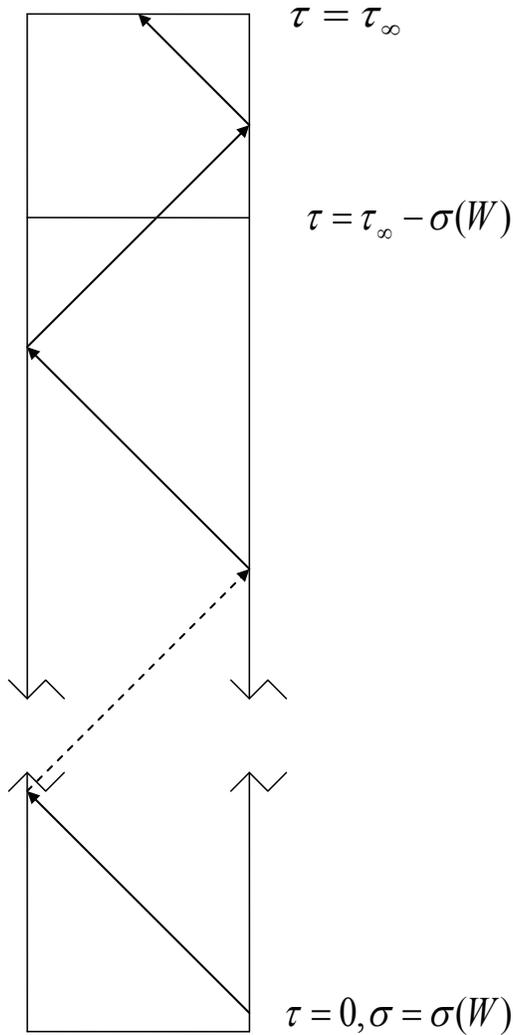}
\caption{Penrose diagram of Linde's proposal: the square at the top represents the usual inflationary era; below it is the pre-inflationary era, with global windings shown. Here $\tau$ is conformal time, $\sigma$ is a conformal radial coordinate, and W refers to the Weeks manifold: see Section 3.}
\end{figure}
The arrows in the diagram represent a signal winding around a compact spatial section. This ``global winding" is only possible because the expansion is relatively slow during the pre-inflationary era; it ceases to be possible precisely when the characteristic inflationary explosive expansion sets in. [The diagram is of course schematic; the behaviour of geodesics, particularly in the negatively curved case, is much more complex than what the diagram suggests.] Notice that the diagram must be substantially taller than it is wide if multiple global windings are to be possible in the pre-inflationary era.

The point of this construction is this. With compact negatively curved spatial sections, the ergodicity of the geodesic flow means that, as Cornish, Spergel, and Starkman have explained \cite{kn:mixing}, global winding implies that multi-point correlations decay exponentially quickly as field modes propagate in this region of spacetime. This effect, known as \emph{chaotic mixing}, can very easily maintain the initial homogeneity in the pre-inflationary era. In short, in Linde's approach, non-trivial topology can temporarily suspend the growth of gravitational entropy, so that the spatial sections are still extremely homogeneous by the end of the pre-inflationary era. This solves the problem: Inflation can begin even though the Universe may by then be three or four orders of magnitude larger than it was initially. A tiny homogeneous region has become a relatively large homogeneous region.

Spatial sections of the form H$^3$/$\Gamma$ are
in fact very natural from a string-theoretic point of view: string
winding ``transforms geometry to topology" exactly when one
compactifies H$^3$ in the way Linde suggests [see
\cite{kn:starr}\cite{kn:silvery}]. From another point of view, the advantages of compact spatial sections in the Hartle-Hawking or ``creation from nothing" approach [see \cite{kn:kieferbook} for a detailed review] have been discussed with various emphases by Zeldovich and Starobinsky
\cite{kn:starobinsky}, by Coule and Martin \cite{kn:martin}, by
Linde \cite{kn:lindetypical}\cite{kn:lindenew}, and by Ooguri, Vafa, and Verlinde \cite{kn:ooguri}. Linde in particular
stresses that if a spacetime is produced by quantum creation
from ``nothing", compact flat or negatively curved spatial sections are
likely to be favoured, because in this case there is no barrier
through which one must tunnel; in short, small worlds are
easier to create\footnote{Note that, in most inflationary approaches, the non-trivial topology is \emph{not} directly
observable \cite{kn:janna}\cite{kn:tavakol} at the present time, or indeed at any
time in the future; see however \cite{kn:lindeopen}\cite{kn:turokopen} for observational issues arising in ``open inflation".}.

Linde's proposal is very appealing, because it offers an extremely natural way of decoupling the scale at the beginning of Inflation from the Planck or string scales. As with all good solutions of deep problems, however, it is not clear that it can actually be made to work. The key point is this. In any cosmological model where the spatial sections are compactified, we are effectively removing all fluctuations larger than a certain size. The well-known result of this is that we can expect the \emph{Casimir effect} to arise
\cite{kn:bytsenko}. The Casimir effect arising cosmologically when
negatively curved spaces are compactified has been studied extensively: see for example
\cite{kn:helio}\cite{kn:mark}.
Typically, the Casimir effect has a local manifestation in terms of
an energy density, which definitely couples to gravity
\cite{kn:fulling}, and which violates the Null Energy
Condition or NEC.

The Casimir effect plays a fundamental role here, because it places several serious obstacles in the path of the ideas of \cite{kn:lindetypical}. Firstly, it forbids the spatial sections of the Universe to be arbitrarily small, and so there \emph{is} a barrier through which the Universe must tunnel. Secondly, Casimir energy tends to \emph{shorten} the Penrose diagram of the pre-inflationary spacetime; we shall explain this in the next section. If this effect is too marked, then of course the purpose of the compactification will have been defeated: if the Penrose diagram ceases to be very tall relative to its width, then causality rules out chaotic mixing.

The third and most dangerous consequence of Casimir-induced NEC violation is that it can very easily render the system unstable. In particular, it is known that a specifically string-theoretic effect can lead to serious instabilities when the NEC is violated. This effect, discovered by Seiberg and Witten \cite{kn:seiberg} and applied to cosmology by Maldacena and Maoz \cite{kn:maoz}, was studied in the case of spatially \emph{flat} compact\footnote{There is an extensive literature on the use of
the Casimir effect in cosmological models with compact flat spatial
sections: see for example
\cite{kn:oddy}\cite{kn:levin}.} pre-inflationary spacetimes in \cite{kn:singularstable}. It was found in that case that the stringy instability due to Casimir violations of the NEC do indeed cause the pre-inflationary system to become unstable very quickly ---$\,$ well before Inflation can begin. We see that the mechanism of \cite{kn:lindetypical} \emph{does not always work}, at least when Inflation is embedded in string theory. This is actually a point in its favour, since it brings us closer to an understanding of how spacetime topology is determined physically: toral topology is ruled out\footnote{Except possibly at the moment of creation, when the topology itself may be ambiguous in string theory: see the Conclusion.}. The question, of course, is whether similar considerations also rule out the hyperbolic case. We shall see that this does happen, for sufficiently large values of the gravitational Casimir coupling.

In summary, then, large values of the Casimir gravitational coupling are fatal to Linde's proposal: they can prevent the Universe from being sufficiently small at its birth, they can prevent chaotic mixing, and they can render the whole system unstable. Linde points out, however, that the Casimir effect is suppressed by supersymmetry [see for example \cite{kn:suppressed}], and he suggests that the weakness of supersymmetry breaking in our Universe may be explained by the fact that small values of the Casimir coupling are evidently favoured by the tunnelling approach to quantum cosmology. This is potentially a very important insight, because a full embedding of Inflation in string theory depends on a precise understanding of supersymmetry breaking.

In order to pursue this idea, we need some quantitative results: how small must the Casimir coupling be in order to avoid the problems discussed above? Answering this question [which is implicitly raised but not investigated in \cite{kn:lindetypical}] is the main objective of the present work.

Clearly we need to be able to fix or constrain the shape of the Penrose diagram. For that, we need to make reasonable assumptions regarding the matter content of the earliest Universe. We assume this to consist of the inflaton, in its potential-dominated state [in which we hope to maintain it by chaotic mixing], together with a [negative] Casimir energy density. The latter is diluted by the slow expansion during the pre-inflationary era; by the time Inflation in the usual sense begins [when, as above, the expansion becomes so fast that circumnavigations of the Universe cease to be possible, and chaotic mixing ceases to be relevant], the Casimir energy has been diluted to negligible levels ---$\,$ assuming of course that the system has not become unstable in the meantime. With these assumptions we can study the question of the \emph{height} of the Penrose diagram in this case, and this is the topic of Section 2.

The question of the \emph{width} of the diagram is complicated by the fact that, while the volume of a compact flat space can be freely prescribed, the volumes of compact negatively curved spaces are fixed by the magnitude of the curvature and by the topology of the space. Furthermore, the size of the smallest possible such space was, until very recently, a matter of conjecture. This basic question has however been settled in a major work due to Gabai et al.
\cite{kn:gabai}\cite{kn:gabaiagain}, who proved that the well-known \emph{Weeks
manifold} is the compactification of H$^3$ with the smallest
possible volume.

If we follow Linde's \cite{kn:lindetypical} argument, that it should be ``easier
to create" small universes than large ones, to its
logical conclusion, then the Weeks manifold is of particular interest. Furthermore, there is growing evidence \cite{kn:gabby}\cite{kn:gabaiagain} for
Thurston's long-standing conjecture that there is a precise
relationship between the combinatorial/topological complexity
of a compact hyperbolic manifold and its volume, and it is
reasonable to argue
\cite{kn:twamley}\cite{kn:gibneg1}\cite{kn:gibneg2} that
quantum-gravitational effects favour low complexity. For these reasons, and for the sake of concreteness, we
shall assume that the spatial sections of the Universe have the topology of the Weeks manifold. This gives us a basis for describing the width of the Penrose diagram. This is the topic of Section 3. Combining these results with those of Section 2, we are in a position to discuss the shape of the Penrose diagram in a quantitative manner. This allows us to estimate the extent to which the Casimir coupling must be suppressed by supersymmetry in order for Linde's suggestion to work.

The danger that Seiberg-Witten instabilities might arise when the spatial sections are negatively curved is the topic of Section 4. The negatively curved case is very different from, and presents several technical difficulties compared with, the flat case. We are nevertheless able to show that spacetimes of this kind can indeed be unstable in the Seiberg-Witten sense; however, this does not occur, as it does in the case of flat spatial sections, for \emph{all} values of the Casimir gravitational coupling: it only occurs when the latter is large enough. Fortunately, the permitted values of the Casimir coupling are compatible with our earlier findings.

We begin with a study of the background spacetime structure and of the way it is deformed when the Casimir effect is taken into account.

\addtocounter{section}{1}
\section* {\large{\textsf{2. The Height of the Penrose Diagram}}}
In this section, we shall introduce a simple explicit spacetime
geometry, arising when the effects of Casimir energy are
superimposed on a spatially compactified version of de Sitter spacetime in the slicing by hyperbolic spatial sections. For the sake of clarity, let us recall the details of the latter.

The [simply connected version of] global de Sitter spacetime with
[in the signature we use here] spacetime curvature 1/L$^2$ is
defined as the locus, in five-dimensional Minkowski spacetime
[signature ($-\;+\;+\;+\;+$)], defined by the equation
\begin{equation}\label{eq:A}
\m{-\; A^2\; + \;W^2 \;+ \;Z^2\; +\; Y^2\; + \;X^2\; =\; L^2}.
\end{equation}
This locus has topology $\bbr\times\,\m{S}^3$, and it can be
parametrized by \emph{global} conformal coordinates
($\eta,\,\chi,\,\theta,\,\phi)$ defined by
\begin{eqnarray} \label{eq:B}
\m{A} & = & \m{L\;cot(\eta)  }                     \nonumber \\
\m{W} & = & \m{L\;cosec(\eta)\;cos(\chi)}                     \nonumber \\
\m{Z} & = & \m{L\;cosec(\eta)\;sin(\chi)\;cos(\theta)}           \nonumber \\
\m{Y} & = & \m{L\;cosec(\eta)\;sin(\chi)\;sin(\theta)\;sin(\phi)}  \nonumber \\
\m{X} & = & \m{L\;cosec(\eta)\;sin(\chi)\;sin(\theta)\;cos(\phi)}.
\end{eqnarray}
Here $\chi,\,\theta,\,\phi$ are the usual coordinates on the
three-sphere, and $\eta$ is angular conformal time, which takes its
values in the interval ($0,\;\pi$). The metric of Global de Sitter
spacetime is then
\begin{equation}\label{eq:C}
\m{g(GdS)\; =\; L^2\,cosec^2(\eta)[ -\; d\eta^2 \; +\; d\chi^2 \;+\;
sin^2(\chi)\{d\theta^2  \;+\; sin^2(\theta)d\phi^2\}}].
\end{equation}
An obvious conformal transformation allows us to extend the range of
$\eta$, so that it takes all values in the closed interval
[$0,\;\pi$]. The Penrose diagram is clearly square [in the case of
\emph{simply connected} spatial sections], since $\chi$ also has
this range. This square diagram rules out global causal contact, since no signal from the antipode of a given point can ever reach it in a finite time. [Recall the discussion around Figure 1: the topmost square, in which circumnavigations are clearly impossible, is described to a good approximation by a metric similar to the one in equation (\ref{eq:C}).]

Now notice that the defining formula (\ref{eq:A}) is invariant under
an exchange, followed by a simultaneous complexification\footnote{It
is convenient to rotate A and W in opposite directions.}, of A and
W; so this transformation cannot change the local geometry.
Therefore, if we perform the exchange and complexify both $\eta$ [to
complexify A and W] and $\chi$ [so as then to avoid complexifying X,
Y, and Z], the resulting coordinates, defined by
\begin{eqnarray} \label{eq:BDS}
\m{A} & = & \m{-\,L\;cosech(\tau)\;cosh(\sigma)  }                     \nonumber \\
\m{W} & = & \m{L\;coth(\tau)}                     \nonumber \\
\m{Z} & = & \m{L\;cosech(\tau)\;sinh(\sigma)\;cos(\theta)}           \nonumber \\
\m{Y} & = & \m{L\;cosech(\tau)\;sinh(\sigma)\;sin(\theta)\;sin(\phi)}  \nonumber \\
\m{X} & = &
\m{L\;cosech(\tau)\;sinh(\sigma)\;sin(\theta)\;cos(\phi)},
\end{eqnarray}
are still coordinates on a spacetime locally identical to de Sitter
spacetime. However, complexification will change the nature of the
coordinates; the periodic coordinates are replaced by coordinates
taking values in an infinite range. Thus the new conformal time
coordinate $\tau$ ranges from zero to infinity, as does the
coordinate $\sigma$ which replaces $\chi$. The effect of this is
actually to \emph{restrict} the domain of these new coordinates:
they cannot cover the entire spacetime, because global de Sitter has
compact spatial sections. [This is analogous to the fact that stereographic coordinates on the sphere cannot cover it completely: they cover the sphere minus a point.]

Comparing the expressions for A in
($\ref{eq:B}$) and ($\ref{eq:BDS}$), we see that $\eta\,> \, \pi/2$
on the domain of these coordinates, and then a comparison of the two
expressions for W shows that
\begin{equation}\label{eq:E}
\eta \;>\;{{\pi}\over{2}}\;+\;\chi.
\end{equation}
We see that the new coordinates actually parametrise only one-eighth
of the full Penrose diagram, the triangular top left-hand corner
extending upwards from the point $\chi$ = 0, $\eta$ = $\pi$/2. Paradoxically, this is precisely what allows the spatial sections of this sub-manifold to be infinite: in order to remain within this corner, they are forced to bend upwards towards future conformal infinity. Clearly the spacetime so obtained is geodesically incomplete, but, by compactifying the spatial sections, we can confine this incompleteness to $\chi$ = 0, $\eta$ = $\pi$/2. As we shall see, the region around this point will eventually be replaced by another geometry when the effects of the Casimir energy are taken into account, and in this way we obtain a complete spacetime.

The ``Spatially Hyperbolic de Sitter" or SHdS metric in
these coordinates is
\begin{equation}\label{eq:H}
\m{g(SHdS)\; =\; L^2\,cosech^2(\tau)[ -\; d\tau^2 \; +\; d\sigma^2
\;+\; sinh^2(\sigma)\{d\theta^2  \;+\; sin^2(\theta)d\phi^2\}}].
\end{equation}
We see at once that this piece of de Sitter spacetime is indeed foliated by
spacelike hypersurfaces of constant \emph{negative} curvature $-1/$L$^2$. One
sees this also if one uses coordinates (t, r, $\theta$, $\phi$)
based on proper time: the same metric is now
\begin{eqnarray}\label{eq:G}
\m{g(SHdS})\; =\;
\m{-\,dt}^2\;+\;\m{sinh^2\big(t/L\big)}\,\Big[\mathrm{dr^2\;
+\;\m{L^2}\, sinh^2(r/L)}\{\mathrm{d}\theta^2 \;+\;
\mathrm{sin}^2(\theta)\,\mathrm{d}\phi^2\}\Big];
\end{eqnarray}
here r = $\sigma$L.

Global de Sitter spacetime, with its spherical spatial sections, can be regarded as the Lorentzian version of a Euclidean sphere. In the same way, Spatially Hyperbolic de Sitter is closely allied to the four-dimensional hyperbolic space H$^4$, which has metric
\begin{equation}\label{eq:W}
\m{g(H^4)\; =\; dt^2\;+\;L^2\,sinh^2\big(t/L\big)\,\Big[d\chi^2\;
+\;sin^2(\chi)\{d\theta^2 \;+\; sin^2(\theta)\,d\phi^2}\}\Big].
\end{equation}
The extrinsic geometry of the t = constant sections here is identical to that of the spatial sections in SHdS. One can in fact obtain g(H$^4$) from g(SHdS) by complexifying t and L and re-labelling r appropriately. This will be useful later.

Unless we compactify, SHdS is spatially infinite ---$\,$ the spatial coordinate $\sigma$ in equation (\ref{eq:H}) ranges from zero to infinity. Similarly,
conformal time is also infinite. That is, $\tau$ runs
from 0 [corresponding to t = $\infty$] to $\infty$ [as t tends to
0].

A spatial compactification of SHdS will, as discussed above, give rise to a Casimir energy. Because it violates various
energy conditions, including the Null Energy Condition [NEC], this has the highly desirable effect of removing
the region near to t = 0 in the pure SHdS geometry, and replacing it with a geometry in which the scale factor is never zero. The upshot is that the range of conformal time is not really infinite, since t = 0 corresponds to infinite conformal time in the original geometry. On the other hand, the compactification itself effectively renders finite the range of $\sigma$. We see, then, that the effect of the spatial compactification is to force the Penrose diagram to be finite in \emph{both} directions. Our task is to determine the precise shape of this diagram.

The general theorem governing the shape of Penrose diagrams in these situations is
the beautiful result due to Gao and Wald
\cite{kn:gaowald} [see
also \cite{kn:gregnull}]:

\bigskip
\noindent \textsf{THEOREM [Gao-Wald]: Let M be a spacetime
satisfying the Einstein equations and the following conditions:}

\noindent \textsf{[a] The Null Energy Condition [NEC] holds.}

\noindent \textsf{[b] M is globally hyperbolic and contains a
compact Cauchy surface.}

\noindent \textsf{[c] M is null geodesically complete and satisfies
the null generic condition.}

\noindent \textsf{Then there exist Cauchy surfaces S$_1$, S$_2$,
with S$_2$ $\subset$ I$^+$[S$_1$], such that, for any p $\in$
I$^+$[S$_2$], one has S$_1$ $\subset$ I$^-$[p].}

\bigskip

Here the null generic condition is the requirement that, along every
null geodesic, there should exist a point where the tangent vector
k$^{\m{a}}$ and the curvature R$_{\m{abcd}}$ satisfy
$\m{k_{[a}\,R_{b]cd[e}\,k_{f]}\,k^c\,k^d \neq 0}$, and I$^+$, I$^-$
denote respectively the chronological future [past] of an event or
set of events; see \cite{kn:waldbook}, Chapter 8. The Null Energy
Condition is the demand that the [full]
stress-energy-momentum tensor should satisfy
\begin{equation}\label{I}
\m{T_{ab}\,n^{a}\,n^{b}\;\geq\;0}
\end{equation}
at all points in spacetime and for all null vectors $\m{n^{a}}$.

In simple language, the Gao-Wald theorem means that a sufficiently
long-lived observer will, under the stated conditions, ultimately be
able to ``see" an entire spatial slice of the spacetime. An
even simpler way of thinking about the theorem is as follows. Take
simply connected global de Sitter spacetime, and note that the
conclusion of this theorem is not true of it. This is because
de Sitter spacetime is so ``special" that it does not actually
satisfy the null generic condition. In fact, because the conformal
diagram is square, global de Sitter spacetime just barely
escapes having a ``fully visible" spatial section. The Gao-Wald theorem
means that, if generic matter satisfying the NEC is introduced into de Sitter spacetime, the effect is to cause the conformal
diagram to become ``\emph{taller}", because in a spacetime with such a diagram there \emph{is} a fully visible
spatial section. Of course,
this affects all parts of the diagram, so even if we decide to cut
off part of the spacetime at a finite time, the remaining part of
the diagram is also stretched vertically when such matter is
introduced\footnote{See \cite{kn:tallandthin} for an explicit example of this.}.
Conversely, the introduction of matter ---$\,$ such as negative Casimir energy ---$\,$ which violates the NEC
will compress every part of the diagram in the vertical direction.
As we mentioned in Section 1, this could possibly interfere with the idea suggested in \cite{kn:lindetypical}.

Let us set up a simple Friedmann model of this system; in doing so,
we are as usual ignoring the back-reaction of the inhomogeneities in
the Casimir energy distribution. [In the case of the Weeks manifold
with which we are concerned here, the relative inhomogeneities of
the Casimir energy tend in any case to be very mild; see
\cite{kn:helio}.] We shall work with the usual FRW spacetime
geometry, with scale factor a(t), where t is proper time, and with
negatively curved spatial sections on which we continue to use the
coordinates (r, $\theta$, $\phi$). [The fact that these sections
have been compactified is not apparent in the form of the local
metric; it is reflected only in the range of r, assuming that we
take this coordinate to be single-valued. See below.]

Various kinds
of physical fields and compactification schemes contribute to the
Casimir energy in various ways and with different signs. If the Casimir energy density is positive,
then the NEC is satisfied and the Gao-Wald theorem shows that the effect is to make the Penrose diagram
taller than it would otherwise have been. The reverse is true in the case of a negative Casimir energy, to which we
now turn. Now we have a new contribution to the energy density, one which is negative and depends on the inverse
fourth power of the scale factor. This is only an approximation to the one-loop
correction of the appropriate effective action, but it is a good approximation
in cases like this where the initial rate of expansion is very slow and the compactification
scale is [supposed to be] much smaller than the horizon radius [of the universal covering spacetime] ---$\;$ see for example \cite{kn:levin}\cite{kn:saharam}.

Thus, the Casimir energy density $\rho_{\m{casimir}}$ is
given by $-\,3\gamma/8\pi$L$^2$a$^4$, where $\gamma$ is a positive dimensionless constant; we are measuring the Casimir energy relative to the positive vacuum energy density
$\rho_{\m{inflaton}}\,=\,+\,3/8\pi$L$^2$, representing the inflaton
in its potential-dominated state. Here L is the typical inflationary
length scale [that is, the Hubble parameter at the end of
Inflation is 1/L, and 3/L$^2$ is the effective cosmological constant], and we are using Planck units. The Friedmann equation is
\begin{eqnarray}\label{eq:J}
\m{L^2\,\dot{a}^2\;=\;{{8\pi}\over{3}}\,L^2\,a^2\Big[\rho_{\m{inflaton}}\,+\,
\rho_{\m{casimir}}\Big]\;+\;1\;=\;{{8\pi}\over{3}}\,L^2\,a^2\Big[{{3}\over{8\pi
L^2}}\;-\;{{3\gamma}\over{8\pi L^2\,a^4}}\Big]\;+\;1.}
\end{eqnarray}

The solutions for the ``Spatially Hyperbolic de Sitter plus Casimir" metric [with
the constant of integration absorbed into the time coordinate] are
\begin{eqnarray}\label{eq:K}
\m{g(SHdS\alpha}) = \m{-\;dt^2 + \Big[
\alpha^2+(1\,+\,2\alpha^2)\,sinh^2(t/L)\Big]\Big[dr^2 + \m{L^2}\,
sinh^2(r/L)}\{\mathrm{d}\theta^2 +
\mathrm{sin}^2(\theta)\,\mathrm{d}\phi^2\}\Big],
\end{eqnarray}
where
\begin{equation}\label{eq:L}
\alpha^2\;=\;\sqrt{{1\over 4}\;+\;\gamma}\;-\;{1\over 2}.
\end{equation}
The metrics are labelled by $\alpha$; of course, $\alpha$ = 0 corresponds to Spatially Hyperbolic de Sitter itself.
Note that these metrics are not singular: $\alpha$ is the smallest possible value of the scale
factor\footnote{\emph{Positive} Casimir energy, by contrast, always leads to a singular metric. One can see this directly for small negative values of $\gamma$, since these correspond to small negative values of $\alpha^2$; from equation (\ref{eq:K}) this implies that the spatial volume vanishes, and the Casimir energy density consequently diverges, for a sufficiently small t.}. This gives the geometric meaning of $\alpha$: if for example
we compactify using the Weeks manifold, then [see
the next section] the minimal volume of a spatial section in this
spacetime is approximately 0.9427$\alpha^3$L$^3$, where L is the
inflationary length scale. Thus $\alpha$ measures the barrier through which the Universe must tunnel from
``nothing". We can also see that $\alpha$ has another physical interpretation: the maximal
magnitude of the Casimir energy density [which is clearly attained
at a unique time, t = 0] is
$|\rho^{\m{max}}_{\m{casimir}}|\,=\,3\gamma/8\pi$L$^2\alpha^4$, and so the total amount of Casimir energy in the Universe initially is
\begin{equation}\label{eq:LL}
\m{|E_{cas}(t = 0)|\;=\;0.9427\alpha^3L^3|\rho^{max}_{casimir}|\;=\;{3\times 0.9427 \, L\over8\pi}\bigg(\alpha\;+\;\alpha^3\bigg)}.
\end{equation}
Thus $\alpha$ can be thought of as a number which parametrises the initial quantity of Casimir energy. For these reasons it is convenient to
work with $\alpha$ rather than the gravitational Casimir coupling $\gamma$ itself. When we need to restore $\gamma$,
we have the simple relation
\begin{equation}\label{eq:LLAMA}
\gamma\,=\,\alpha^2\,+\,\alpha^4.
\end{equation}

The formal
``total equation of state parameter" [the ratio of total pressure to
total energy density] is given in this cosmology by
\begin{equation}\label{eq:LLL}
\m{w_{tot}\;=\;{-\,(1\;+\;\gamma/3a^4)\over (1\;-\;\gamma/a^4)}},
\end{equation}
where a(t) is the scale factor. Notice that, since $\gamma^{1/4}$ is, by (\ref{eq:LLAMA}), slightly larger than $\alpha$, w takes on
arbitrarily large negative values for small values of t. [It rapidly approaches
$-\,1$ as t increases, however.]

It will be useful for us to write our metrics in terms of
[dimensionless] conformal time, $\tau$, which is given by
\begin{equation}\label{eq:M}
\m{\tau\;=\;{1\over \alpha}\int_0^{t/L}{dt/L\over
\sqrt{1\;+\;[2\;+\;(1/\alpha^2)]sinh^2(t/L)}}\;=\;{- i \over
\alpha}\,F\Bigg({i t \over L}\;,\;\sqrt{2\;+\;(1/\alpha^2)}\Bigg)}.
\end{equation}
Here F($\phi \;,\;$ k) is the incomplete elliptic integral of the
first kind \cite{kn:abramo}, with Jacobi amplitude $\phi$ and
elliptic modulus k; in this case it has been evaluated along the
imaginary axis.

Inverting the elliptic integral we can express t in terms of the
amplitude:
\begin{equation}\label{eq:N}
\m{it/L\;=\;am\Big(i\alpha\tau \;,\;\sqrt{2\;+\;(1/\alpha^2)}\Big)}.
\end{equation}
Taking the sine of both sides we find that
\begin{equation}\label{eq:O}
\m{i\,sinh(t/L)\;=\;sn\Big(i\alpha\tau\;,\;\sqrt{2\;+\;(1/\alpha^2)}\Big)},
\end{equation}
where sn(u, k) is one of the classical Jacobi elliptic functions.
Using the formulae for complex arguments of elliptic functions given
on page 592 of \cite{kn:abramo}, one can express the right side as a
function of a real variable; substituting the result for sinh(t/L)
in equation (\ref{eq:K}) we obtain finally [with $\sigma$ = r/L]
\begin{eqnarray}\label{eq:P}
\m{g(SHdS\alpha}) &=& \m{ L^2\,\Bigg[
\alpha^2+(1\,+\,2\alpha^2)\,{sn^2\Big(\alpha\tau\;,\;i\,\sqrt{1\;+\;(1/\alpha^2)}\Big)\over
cn^2\Big(\alpha\tau\;,\;i\,\sqrt{1\;+\;(1/\alpha^2)}\Big)}\Bigg]\;}
\nonumber \\
& & \times\;\m{\Big[ -\; d\tau^2 \; +\; d\sigma^2 \;+\;
sinh^2(\sigma)\{d\theta^2 \;+\; sin^2(\theta)d\phi^2\}\Big]. }
\end{eqnarray}
Here cn(u, k) is another of the Jacobi elliptic functions. This
metric is to be compared with the Spatially Hyperbolic de Sitter metric given in
equation (\ref{eq:H}); the metric here is conformally the same as
that metric along future infinity; it is asymptotically de Sitter.

The geometry here differs from that of Spatially Hyperbolic de Sitter spacetime in an
important way, however: the extent of conformal time is \emph{not}
infinite. Its extent is instead given by setting the elliptic
function
$\m{cn\Big(\alpha\tau\;,\;i\,\sqrt{1\;+\;(1/\alpha^2)}\Big)}$ equal
to zero. The zeros of this function are given [\cite{kn:abramo},
page 590] by
\begin{equation}\label{eq:Q}
\m{cn(K(k), k)\;=\;0,}
\end{equation}
where K(k) is the complete elliptic function of the first
kind. Thus $\tau$ has a formal range between
\begin{equation}\label{eq:R}
\m{\pm\,\tau_{\infty}(\alpha)\;=\;\pm\,{1\over
\alpha}\,K\Big(i\,\sqrt{1\;+\;(1/\alpha^2)}\Big).}
\end{equation}

\begin{figure}[!h]
\centering
\includegraphics[width=0.7\textwidth]{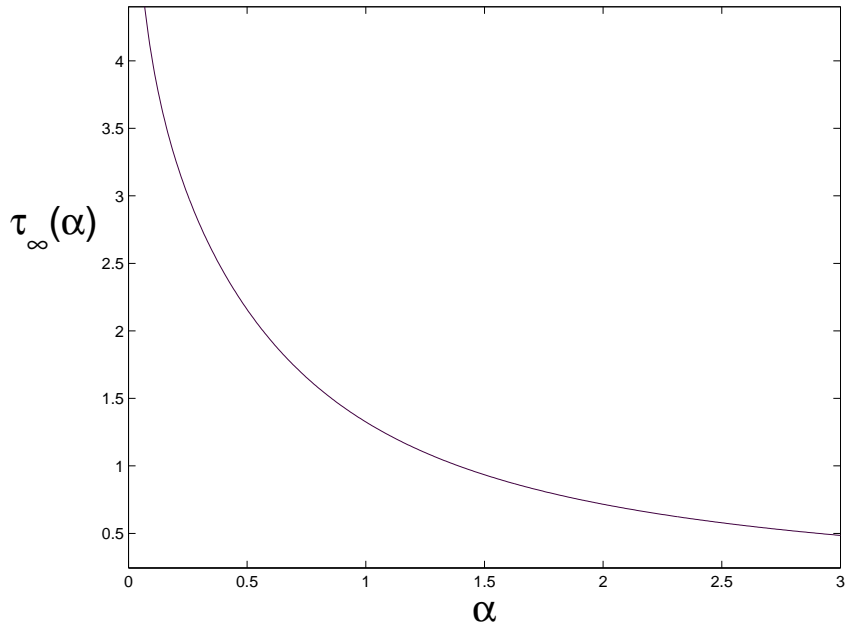}
\caption{Conformal Time to Infinity as a Function of $\alpha$.}
\end{figure}

Using the relevant formula from page 593 of \cite{kn:abramo} one can
express this in terms of real variables:
\begin{equation}\label{eq:S}
\m{\pm\,\tau_{\infty}(\alpha)\;=\;\pm {1\over
\sqrt{1\;+\;2\alpha^2}}\,K\,\Bigg(\sqrt{{1\;+\;\alpha^2\over
1\;+\;2\alpha^2}}\Bigg).}
\end{equation}
However, the negative value here corresponds to proper time t
tending to $-\,\infty$, which is not correct; it should instead be
replaced by the conformal time at which the Universe tunnels from ``nothing". This corresponds to the
spatial hypersurface of zero extrinsic curvature, which is located at t = $\tau$ = 0. Thus $\tau_{\infty}(\alpha)$
is exactly what we are seeking, the full height of the Penrose diagram [from the creation until the end of Inflation].

Since the function K(k) diverges as k tends to unity
[it increases monotonically from a value of $\pi$/2 at k = 0, and in
particular is finite at $1/\sqrt{2}$ ---$\,$ see \cite{kn:abramo},
page 592], $\tau_{\infty}(\alpha)$ can be arbitrarily large if
$\alpha$ is very small, or arbitrarily small if $\alpha$ is large;
see Figure 2.

The graph of $\tau_{\infty}(\alpha)$ is a very graphic illustration of the Gao-Wald theorem: as the NEC-violating effect becomes stronger, the Penrose diagram is compressed vertically. In fact, we see that, by adjusting
$\alpha$, we can obtain \emph{any} prescribed shape for the Penrose diagram. We also see that $\alpha$ has to be quite small if the Penrose diagram is to be tall, as we wish. In order to be more precise, we need some information on the width of the diagram. We now turn to this.

\addtocounter{section}{1}
\section* {\large{\textsf{3. The Width of the Penrose Diagram}}}
It was shown long ago by Thurston \cite{kn:thursty} that there
exists a decomposition of hyperbolic three-space H$^3$ into
identical pieces of minimal volume ---$\;$ that is, that there is a
minimal-volume compactification. It was long conjectured, and
finally proved by Gabai et al. \cite{kn:gabai}\cite{kn:gabaiagain}, that this
distinguished decomposition corresponds to the \emph{Weeks
manifold}\footnote{A good description of the Weeks manifold, with
illustrations, may be found in \cite{kn:inoue}; see also
\cite{kn:gomero}.}, \textbf{W}. For this manifold, the fundamental
domain can be represented as a certain hyperbolic polyhedron with 18
faces. The volume is $\approx$ 0.9427$\times$$\lambda$$^3$, where
$\lambda$ is the curvature radius. Our hypothesis is that this distinguished
hyperbolic compactification is the one realised in Nature.

In any compact manifold M, the \emph{injectivity radius} I(M, p) at
a point p in M is defined as the maximal radius of a sphere centred
at p which does not self-intersect. [That is, the maximal radius
such that the exponential map is injective.] This is the radius beyond which outward directed
geodesics begin to return towards their origin, in at least one direction.
For a sphere or a
torus, this quantity is actually independent of the point p, but
this is not so for compact hyperbolic manifolds, for which
the boundary of a fundamental domain is much more irregular. That
is, the injectivity radius is a function of position on a compact
hyperbolic manifold.

The range of sizes of spheres which can be contained in a compact
hyperbolic space can be surveyed as follows. For each such space one
can define an \emph{injectivity distribution}, a function introduced
by Weeks \cite{kn:weeks} and defined as follows. Let dV/V be the
fraction of the volume of M containing points p with I(M, p)
[measured in units of $\lambda$] lying between the values x and x +
dx. Then the injectivity distribution is the function on the real
line defined by
\begin{equation}\label{eq:T}
\m{ID(M; x)\;=\;(dV/V)/dx.}
\end{equation}
That is, ID(M; x) measures the rate at which the fractional volume
containing points with a given injectivity radius changes with
increasing injectivity radius; integrating it between selected
values of x gives the fraction of the volume of M containing points
with injectivity radii between those values. The curve representing
ID(M; x) [which need not be a continuous function] intersects the x
axis at two points. The smaller of these two values signals the
radius at which it becomes possible for a sphere to self-intersect [by choosing the location of
its centre appropriately];
the larger signals the radius beyond which this must happen.

The functions ID(M; x) are given in approximate form for ten
low-volume hyperbolic spaces in \cite{kn:weeks}. In particular, for
the Weeks manifold \textbf{W} [``Manifold 1" in \cite{kn:weeks}], the injectivity distribution
ID(\textbf{W}; x) is a function which has support on an interval
extending roughly from 0.292 to 0.519. Thus, a sphere of conformal radius less than
0.292 can be located anywhere in \textbf{W} without danger of
intersecting itself; but a sphere of conformal radius larger than
0.519 would have to do so, no matter where it might be located.
[Note that this last quantity varies between roughly
0.5 and 0.6 for the ten low-volume manifolds examined in
\cite{kn:weeks}; however it can be substantially larger than this
for other well-known compact hyperbolic manifolds; it is
approximately 0.996 for the Seifert-Weber space, the most easily
visualised compact hyperbolic manifold \cite{kn:thurston}.]

The fact that the injectivity radius varies with location in \textbf{W} means that there is no single Penrose diagram for the spacetimes we have been discussing. If we define the width of the diagram to mean the maximal value of the conformal radial coordinate $\sigma$ such that the coordinate is single-valued ---$\,$ this is how one defines it in the case of global de Sitter ---$\,$ then one obtains a ``position-dependent Penrose diagram". Once we go a little beyond the injectivity radius at a given point, we will find geodesics returning to a neighbourhood of that point, coming from many directions. Thus chaotic mixing may be possible near to that point. Since chaotic mixing depends on having a ``tall" diagram like the one in Figure 1, having a position-dependent diagram means that, in theory, one might have chaotic mixing in some regions of space and not in others. This might be acceptable ---$\,$ after all, we only need Inflation to begin in \emph{some} region ---$\,$ but if we proceed in that way we will be obliged to estimate the effects of signals propagating into the ``mixed" region from ``unmixed" regions\footnote{Issues of this sort were discussed in the case where the NEC is \emph{not} violated and where the spatial topology \emph{is} trivial in \cite{kn:trodden}.}. For the sake of simplicity, and because the range of injectivity radii for \textbf{W} is, as one sees from the graph of its injectivity distribution, not very wide, we shall proceed under the assumption that chaotic mixing should occur everywhere. The width of the diagram in Figure 1, which we denote [with reference to the conformal coordinate $\sigma$] by $\sigma$(\textbf{W}), is then slightly more than the maximal injectivity radius, 0.519. For definiteness we shall take $\sigma$(\textbf{W}) $\approx$ 0.52. We shall abuse terminology and speak of this as the width of ``the" Penrose diagram.

We are now in a position to discuss the shape of Figure 1 in a quantitative way.
Since chaotic mixing is exponentially efficient in removing any inhomogeneities which may develop, we do not need a very large number of circumnavigations to be possible during the pre-inflationary era. Let us assume that five to ten such complete windings are possible; we shall show that this assumption leads to reasonable consequences.

In this case, the lower section of Figure 1 needs to be ten [or twenty] times as high as it is wide, and the whole diagram needs to have a height of 11$\times$$\sigma$(\textbf{W}) $\approx$ 5.72 [or 21$\times$$\sigma$(\textbf{W}) $\approx$ 10.92]. The corresponding values of $\alpha$ then lie, according to equation (\ref{eq:S}), in the range
\begin{equation}\label{eq:STROLCH}
0.00010235 \;\;\leq \;\;\alpha \;\; \leq \;\;0.01853,
\end{equation}
the larger of the two bounds being the one that corresponds to five windings. During the time that these five [ten] windings occur, the scale factor increases by a factor of about 310 [56,000]. Thus, with the assumption we have made regarding the number of windings, Inflation begins when the Universe is roughly 500 to 50,000 times larger than it was at the moment of creation; in other words, this is the extent to which the size of the Universe at the beginning of Inflation has been decoupled from [say] the string length scale. These numbers are reasonable, and in rough agreement with our objectives; they suggest that our assumptions too are reasonable.

Notice that the amount by which the scale factor increases during the pre-inflationary era increases with dramatic speed as the number of windings increases. It follows that if $\alpha$ is made too small [corresponding to a very tall Penrose diagram, and many windings], then the size of the Universe at the inception of Inflation will be unreasonably large. This is why we obtain a \emph{lower} bound on $\alpha$ as well as the expected upper bound. In short, one cannot allow for fewer than five windings, or much more than ten.

Recall that the value of the scale factor at t = 0 is just $\alpha$. Therefore the volume of the Universe when it is created is constrained by (\ref{eq:STROLCH}) according to
\begin{equation}\label{eq:VALDIVIA}
\m{1.011 \times 10^{-12}\,L^3 \;\;\leq \;\;0.9427 \alpha^3 L^3  \;\; \leq \;\;5.998 \times 10^{-6} L^3},
\end{equation}
where we recall that 1/L is the inflationary Hubble parameter. Clearly the barrier through which the Universe must tunnel is very small. Thus, the problem of having to tunnel through a  barrier will be solved automatically if the values of $\alpha$ we need can be achieved.

The corresponding values of the gravitational Casimir coupling are [from (\ref{eq:LLAMA})]
\begin{equation}\label{eq:STROLCHEN}
1.048 \times 10^{-8} \;\;\leq \;\;\gamma \;\;\leq \;\;3.435 \times 10^{-4}.
\end{equation}

Since supersymmetry suppresses the Casimir effect, and yet is broken, one expects a complete theory of supersymmetry breaking to give rise to just such inequalities, constraining the coupling to be small and to be bounded away from zero. This is in agreement with \cite{kn:lindetypical}. It would of course be extremely satisfactory if such a theory could naturally reproduce the above constraints on $\gamma$; conversely one would have to regard the proposal as being falsified if the computed value lies far outside this range.

In summary, then, $\gamma$ must be very strongly constrained if the programme is to work; but the values obtained are not unreasonable if supersymmetry is weakly broken. However, the whole construction is based on a violation of the NEC, which very often leads to instabilities. It is essential therefore to verify that the values we have obtained do not lead to such difficulties.

\addtocounter{section}{1}
\section* {\large{\textsf{4. Stability Despite NEC Violation}}}
The status of the NEC has been much debated of late, from various
points of view [see for example
\cite{kn:od}\cite{kn:brandy}\cite{kn:crem} and references therein].
NEC violation may or may not be acceptable in cosmology, but it is certainly the case that there are many
circumstances in which it leads to major problems [due to ghosts and
gradient energies of the wrong sign
\cite{kn:dubovsky}\cite{kn:kal}]. Exceptions do arise, however, in
strictly quantum, non-local systems ---$\,$ such as those giving rise to the
Casimir effect \cite{kn:nimah}. Arkani-Hamed et al. explain in detail how
this effect is able to avoid the usual perturbative instabilities which otherwise
rule out negative energy densities.

However, while Casimir energy may be acceptable in perturbative physics, it is not
clear that it is innocuous non-perturbatively. String theory provides a concrete
context for investigating this question. In fact, a particular kind of ``stringy" instability 
discovered by Seiberg and Witten\footnote{\emph{Seiberg-Witten instability} refers to the uncontrolled
nucleation of branes in spacetimes with Euclidean versions of a particular kind; it depends on a delicate interplay between the growth of
volumes and surface areas in asymptotically hyperbolic Euclidean
geometries. [A Riemannian manifold is said to be
\emph{asymptotically hyperbolic} if it has a well-defined conformal
boundary. The methods introduced by Seiberg and Witten apply to any
spacetime with an asymptotically hyperbolic Euclidean version.]} \cite{kn:seiberg} is directly relevant here. This effect involves the pair-production of branes\footnote{In order to avoid confusion, we stress that the branes we discuss here are pair-produced
\emph{inside} a given spacetime; there is no connection with braneworld cosmologies.} in a non-perturbative manner; thus it would not be detected in any approach based on perturbative
physics or point-like objects. Furthermore, it is known \cite{kn:unstable} that this particular form
of instability often arises when energy densities are negative. The study of this effect
in \emph{cosmology} was pioneered by Maldacena and Maoz \cite{kn:maoz}, who investigated its consequences
for a particular family of Bang/Crunch
cosmologies by transferring the problem to the Euclidean domain. Other explicit examples of systems which are unstable in this sense were
discussed by Witten and Yau \cite{kn:wittenyau}; see also
\cite{kn:porrati}\cite{kn:tallandthin}. Requiring that Seiberg-Witten instability be absent yields a useful
constraint in many cases: see for example
\cite{kn:conspiracy}\cite{kn:marealle}. Here we can use the methods outlined in Section
2, above, to find the appropriate Euclidean version of Spatially Compactified de Sitter spacetime and variants of it
which contain negative Casimir energy.

While no completely general criterion for the occurrence of
Seiberg-Witten instability is known, one can proceed in simple cases by computing the brane action directly.
The brane action [in the critical case of BPS branes] is readily computed from the area and volume of the brane: see for example \cite{kn:porrati} for the details. The area contributes positively, but the volume negatively, to the action; thus the volume must not grow too rapidly relative to the area if the action is to remain positive. In the case of four-dimensional hyperbolic space H$^4$ [with metric (\ref{eq:W}), discussed earlier], the brane action for a brane of tension $\Theta$ is
\begin{equation}\label{eq:X}
\m{S[H^4](\Theta,
\,L\,;\;t)\;=\;2\pi^2\,\Theta\,L^3\,\Big[sinh^3(t/L)\;-\;{{1}\over{4}}\,cosh(3t/L)\;+\;{{9}\over{4}}\,cosh(t/L)\;-\;2\,
\Big].}
\end{equation}
Note that the negative second term here is actually \emph{larger} in
magnitude than the first term for small values of t, underlining the
fact that the positivity of the action is somewhat precarious even
here, in the case of undisturbed ``pure" hyperbolic space.
Nevertheless, one can verify that [because of the presence of the
third term, which is negligible at large distances] this function,
which obviously vanishes at t = 0, is monotonically increasing, and
hence is everywhere non-negative; thus pure hyperbolic space is itself completely stable in the Seiberg-Witten sense.

When this geometry is deformed by inserting matter, however, it is not at all clear that the action remains positive everywhere; and Maldacena and Maoz \cite{kn:maoz} found in several explicit examples that in fact it need not. To investigate this in our case, recall from Section 2 above that the metric
given in (\ref{eq:W}) is precisely the asymptotically hyperbolic
Euclidean version of the Spatially Hyperbolic de Sitter [SHdS] metric (\ref{eq:G}): one
obtains (\ref{eq:W}) from (\ref{eq:G}) simply by complexifying t and
L and re-labelling r. Performing this same complexification on the metric in
(\ref{eq:K}), we obtain the asymptotically hyperbolic Euclidean version,
with metric
\begin{eqnarray}\label{eq:Y}
\m{g(ESHdS\alpha)} &=& \m{dt^2} \;+\;  \m{L^2\,\Big[
\alpha^2\;+\;(1\;+\;2\alpha^2)\,sinh^2(t/L)\Big]}
\nonumber \\
& & \phantom{aaaaaaaaaa} \times\;\m{\Big[d\chi^2 \;+\;
sin^2(\chi)}\{\mathrm{d}\theta^2 \;+\;
\mathrm{sin}^2(\theta)\,\mathrm{d}\phi^2\}\Big],
\end{eqnarray}
where we truncate at t = 0, which represents the Universe being created from ``nothing".

In this geometry, the brane action for
t $\geq$ 0 is
\begin{eqnarray}\label{eq:Z}
\m{S[ESHdS\alpha](\Theta, \,L\,;\;t)} &=&
\m{2\pi^2\,\Theta\,L^3\,\Bigg[\,\Big(
\,\alpha^2\;+\;(1\;+\;2\alpha^2)\,sinh^2(t/L)\Big)^{3/2}\;} \nonumber \\
                        & & \phantom{aaa} - \;\m{{{3}\over{L}}\,\int_{0}^t\,\Big(
\,\alpha^2\;+\;(1\;+\;2\alpha^2)\,sinh^2(u/L)\Big)^{3/2}du\Bigg]},
\end{eqnarray}
where $\Theta$ is the tension, as in equation (\ref{eq:X}). As the
metric here is asymptotically indistinguishable from that of the
pure hyperbolic space discussed above, this function is certainly
positive at large t. The problem is to understand what happens at
small values of t, where the NEC violation is most intense.

In fact, from a physical point of view we can see that there is indeed a serious danger here. The programme we are discussing in this work is based on the possibility of having a pre-inflationary era during which the expansion of the Universe is slow [by inflationary standards], so that signals can be exchanged on scales such that the topological non-triviality of the spatial sections allows chaotic mixing to operate. But this slow expansion means that the volume of a brane propagating away from the section at t = 0 will grow, while the area hardly changes. For generic values of the parameters, this will certainly lead to the brane action becoming negative. In other words, the very geometry which allows the mechanism to work may doom the system to instability. Fortunately, we are not interested in generic values of the Casimir parameter $\gamma$: we are interested in values satisfying (\ref{eq:STROLCHEN}), above.

The derivative of the action with respect to t can be expressed,
after a straightforward calculation, as
\begin{eqnarray}\label{eq:AA}
\m{{dS[ESHdS\alpha](\Theta, \,L\,;\;t)\over dt}} &=&
\m{3\pi^2\,\Theta\,L^2\,\Big[
\,1\;-\;(1\;+\;2\alpha^2)\,e^{-\,2t/L}\Big]} \nonumber \\
                        & & \phantom{aaaaaaaa} \times\;\m{\Big[
\,\alpha^2\;+\;(1\;+\;2\alpha^2)\,sinh^2(t/L)\Big]^{1/2}}.
\end{eqnarray}

We begin by noting
that the initial value of the action is not zero for $\alpha > 0\,$:
it is equal to the positive value
$\m{2\pi^2\,\Theta\,L^3}\,\alpha^3$. Furthermore, one can show that
the second derivative is positive everywhere. On the other hand, we
see at once that the slope of the graph of the action is
\emph{negative} at t = 0; it is equal to $\m{-\,6
\pi^2\,\Theta\,L^2}\,\alpha^3$. The action function is \emph{not}
monotonically increasing, as it is in the case of pure hyperbolic
space; as expected on physical grounds, there is a real possibility that the action
could become negative. It can be shown that this initial decrease of
the brane action is due to the fact that the Casimir effect
violates the NEC. We see that NEC violation tends to induce
Seiberg-Witten instability, as claimed.

The graph of the action reaches a unique minimum [as can be seen
from (\ref{eq:AA})] at a positive value of t, namely t =
(ln($\sqrt{1 + 2\alpha^2}$))L. The system will be stable in the
Seiberg-Witten sense provided that the action is non-negative at
this point. That is, if we define a number $\Xi_{\alpha}$, depending
only on $\alpha$, by
\begin{eqnarray}\label{eq:BB}
\m{\Xi_{\alpha}\;=\;{S[ESHdS\alpha](\Theta,
\,L\,;\;(ln(\sqrt{1 + 2\alpha^2}))L)\over 2\pi^2\Theta L^3}},
\end{eqnarray}
then the system is non-perturbatively stable if and only if
$\Xi_{\alpha}\,\geq\,0$.

In fact, a numerical investigation shows that the action \emph{does}
become negative if $\alpha$ is sufficiently large, showing that
Seiberg-Witten instability is a possibility here. For example, if we take $\alpha$ = 5, then
it is clear from Figure 3 that the system will be unstable in the
Seiberg-Witten sense. [In this and in the
subsequent diagrams, the t-axis has units given by L; the units on
the vertical axis are given by $2\pi^2\,\Theta\,$L$^3$.]

\begin{figure}[!h]
\centering
\includegraphics[width=0.7\textwidth]{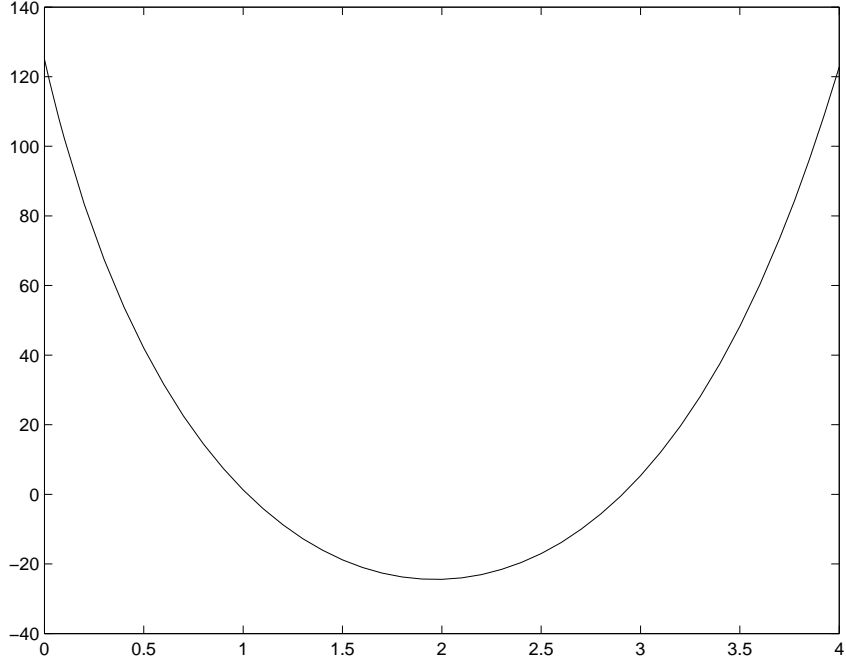}
\caption{Brane Action, $\alpha$ = 5.}
\end{figure}
If we think of $\Xi_{\alpha}$ as a function of $\alpha$, then we see
that this function is already negative at $\alpha$ = 5, and in fact
it becomes steadily more negative as $\alpha$ increases beyond 5. We
know [from equation (\ref{eq:X})] that $\Xi_{\alpha}$ is zero at $\alpha$ = 0, and it
would be perfectly reasonable to expect that it is negative for
\emph{all} positive $\alpha$; this would mean that NEC violation
leads to instability in all cases, which would not be at all
unexpected.
\begin{figure}[!h]
\centering
\includegraphics[width=0.7\textwidth]{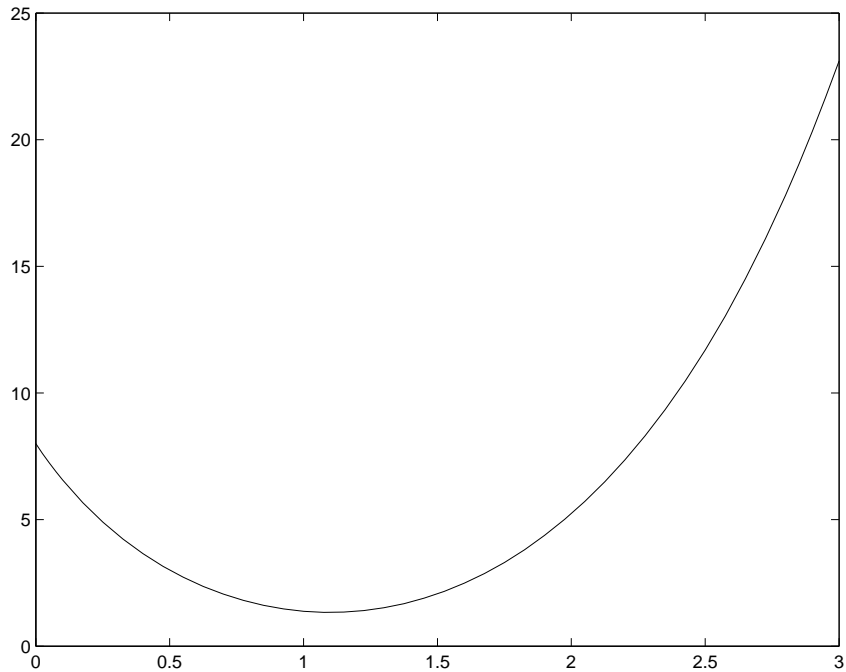}
\caption{Brane Action, $\alpha$ = 2.}
\end{figure}
Remarkably, however, numerical experiments show that this is
not the case: as a function of $\alpha$, $\Xi_{\alpha}$ is
actually positive for a brief interval near to $\alpha$ = 0. For
example, it is positive at $\alpha$ = 2: see Figure 4. Further
experimentation shows that $\Xi_{\alpha}$ is approximately zero at
around $\alpha$ = 2.88, as shown in Figure 5.
\begin{figure}[!h]
\centering
\includegraphics[width=0.7\textwidth]{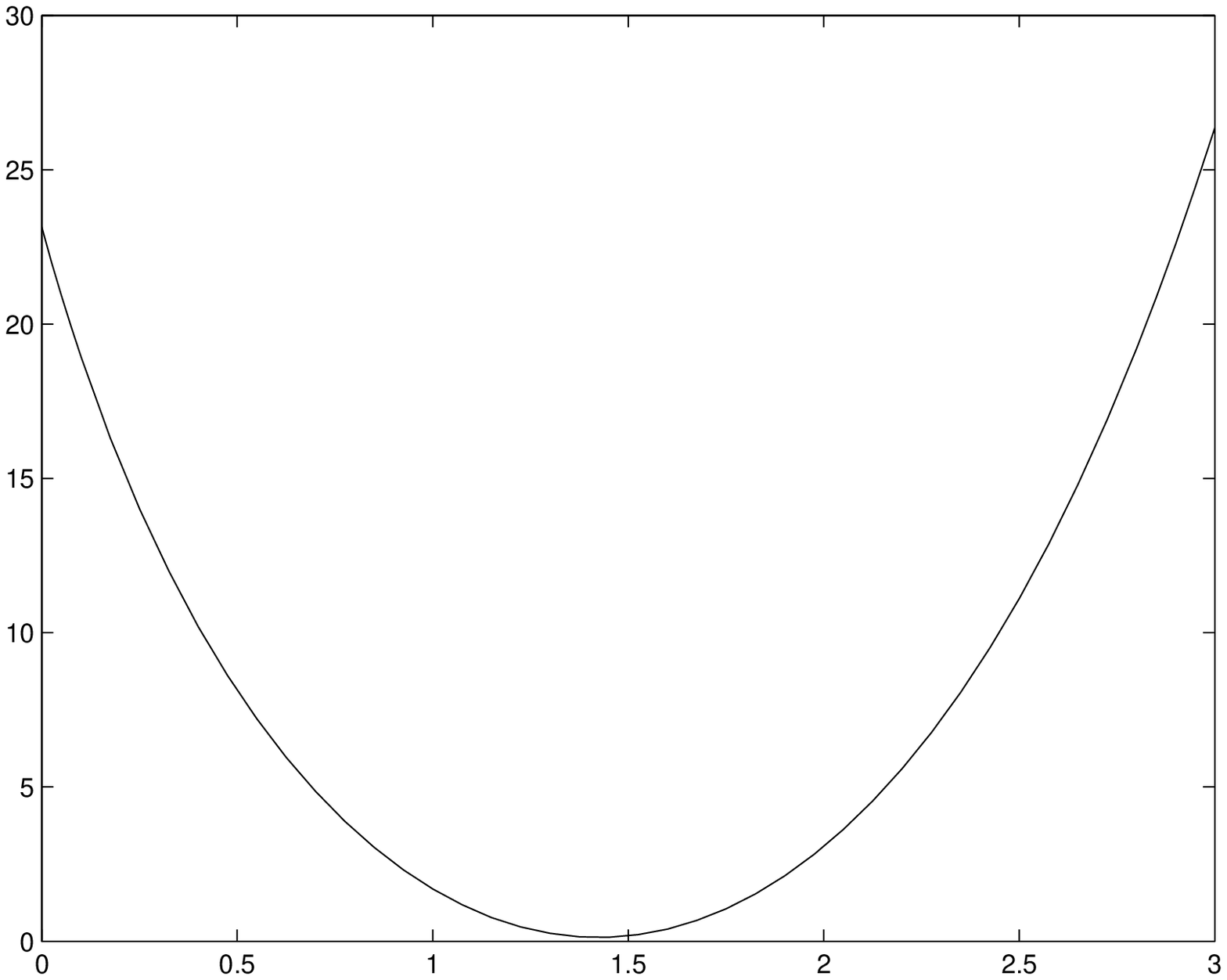}
\caption{Brane Action, $\alpha$ = 2.88.}
\end{figure}
One finds that $\Xi_{\alpha}$ is always negative beyond this value, but that it is always positive for values of $\alpha$ strictly between 0 and 2.88.
Thus, despite the NEC violation associated with the Casimir effect,
the system is actually non-perturbatively \emph{stable} for sufficiently small values of $\alpha$.

Of course, all of these values for $\alpha$ are far larger than those in (\ref{eq:STROLCH}); that is, for the values of $\alpha$ in (\ref{eq:STROLCH}), the action is certainly positive everywhere. Indeed,
from Figure 2 we can see that if $\alpha$ is about 2.88, then, with $\sigma$(\textbf{W}) $\approx$ 0.52
as above, the Penrose diagram will be almost exactly square, which would mean that there is no pre-inflationary era
at all. In other words, the system becomes unstable precisely in the situation where the ideas of \cite{kn:lindetypical} would not work in any case.

In summary, then, we can say that while the spacetime with metric given in (\ref{eq:K}) is unstable in string theory
for \emph{generic} values of $\alpha$, it is completely stable in the cases which arise in Linde's theory.

\addtocounter{section}{1}
\section* {\large{\textsf{5. Conclusion: The Role of Topology}}}
Without Inflation, the problem of understanding the earliest Universe is almost insurmountable: we
would have to understand the initial conditions of all of the constituents, separately. Inflation reduces
this problem to the single puzzle of explaining the origin of the inflationary initial conditions.

Linde has proposed a theory of inflationary initial conditions in the [very common] case where it is necessary for Inflation to begin in a Universe which is already quite large relative to the fundamental scale. It is based on the idea of chaotic mixing due to the non-trivial spatial topology of a universe created from ``nothing". This is an extremely attractive proposal, which, as we have seen, leads to concrete quantitative predictions regarding the Casimir coupling to gravity.
It also leads to a very specific and almost complete picture of the earliest Universe: it is a spacetime with spatial sections isometric to a compact hyperbolic manifold [probably the Weeks manifold] and with a Penrose diagram pictured in Figure 1, with a height about 10 or 20 times its width.

The picture is ``almost" complete because it does not explain the origin of the ultra-low gravitational entropy at the moment when the Universe is created from ``nothing". The generic geometry on a manifold with, for example, the topology of the Weeks manifold, certainly does not have constant negative curvature. That is, the extreme local homogeneity of the spatial section at the time of creation, which we have \emph{assumed} here, needs to be explained.

Of the current theories attempting to explain this basic observation, several \cite{kn:carroll}\cite{kn:mersini} are based on the nucleation of ``baby universes", which are difficult to reconcile with the idea of non-trivial spatial topology; others \cite{kn:bojo}\cite{kn:greene} seem to have no connection with spatial topology, and may have significant fine-tuning problems. Probably the approach which is most nearly compatible with Linde's ideas is the one advanced in \cite{kn:arrow}, where the ``specialness" of cosmic initial conditions is explained in a completely non-dynamical and ``topological" way. The idea is to combine the general-relativistic constraint equations with deep geometric theorems on the space of all possible metrics on compact three-dimensional manifolds of given topology, leading to the conclusion that extreme initial homogeneity is demanded by the internal mathematical consistency of the theory.

Unfortunately, the topology assumed in \cite{kn:arrow} is that of a torus, and the same idea definitely does not work when applied directly to the negatively curved case. However, in \cite{kn:silvery} it is shown that there is a string-theoretic description of  physics on a Riemann surface in terms of its Jacobian torus, and it is argued that a similar duality may hold in higher dimensions. One might hope to be able eventually to show that this duality is operative at the moment of creation from ``nothing"; constraints imposed by the toral topology on one side of the duality, as in \cite{kn:arrow}, might then have an analogous interpretation on the hyperbolic side. This would mean that both the initial homogeneity, as well as its preservation throughout the pre-inflationary era, are explained topologically.

\addtocounter{section}{1}
\section*{\large{\textsf{Acknowledgements}}}
The author is very grateful to Prof Soon Wanmei for help with
numerical work.

\end{document}